\documentclass[pdflatex,sn-mathphys]{sn-jnl}

\jyear{2022}%

\begin{document}

\title[Survival Analysis of the Non-autonomous Ornstein-Uhlenbeck Process]{Analytical Survival Analysis of the Non-autonomous Ornstein-Uhlenbeck Process}


\author[1]{\fnm{L.T.} \sur{Giorgini}}\email{ludovico.giorgini@su.se}

\author[2]{\fnm{W.} \sur{Moon}}\email{woosok.moon@gmail.com}

\author[1,3]{\fnm{J.S.} \sur{Wettlaufer}}\email{john.wettlaufer@yale.edu}

\affil[1]{\orgname{Nordita, Royal Institute of Technology and Stockholm University}, \orgaddress{\city{Stockholm}, \postcode{106 91}, \country{Sweden}}}

\affil[2]{\orgdiv{Department of Environmental Atmospheric Sciences}, \orgname{Pukyong National University}, \orgaddress{\street{45 Yongsoro}, \city{Busan}, \postcode{48513}, \country{South Korea}}}

\affil[3]{\orgname{Yale University}, \orgaddress{\city{New Haven}, \postcode{06520}, \state{CT}, \country{USA}}}


\abstract{The survival probability for a periodic non-autonomous Ornstein-Uhlenbeck process is calculated analytically using two different methods. The first uses an asymptotic approach.  We treat
the associated Kolmogorov Backward Equation with an absorbing boundary by dividing the domain into an interior region, centered around the origin,
and a ``boundary layer'' near the absorbing boundary.  In each region we determine the leading-order analytical solutions, and construct a uniformly valid solution over the entire domain using asymptotic matching. In the second method we examine the integral relationship between the probability density function and the mean first passage time probability density function.  These allow us to determine approximate analytical forms for the exit rate. The validity of the solutions derived from both methods is assessed numerically, and we find the asymptotic method to be superior.}


\keywords{Survival probability, Non-autonomous Ornstein–Uhlenbeck Process, Fokker–Planck equation, Asymptotics}



\maketitle

\section{Introduction}\label{sec1}

A non-autonomous Ornstein-Uhlenbeck (OU) stochastic process describes the evolution of a random variable under the influence of a time-dependent mean-reverting force, and a random source of noise 
\cite{springer53, gordina2020ornstein,digitalcommons55,Ovidio2019}. It has been widely used to model phenomena in physics, biology, finance, and engineering \cite[e.g.,][]{benth2007non,zapranis2008modelling,jahn2011motoneuron,oksendal2013stochastic,keyes2023stochastic,giorgini2022non,giorgini2023thermodynamic}, wherein it provides a simple representation of variability, and allows probabilistic predictions.
One of the important features of OU processes is the survival probability, which is the probability that the random variable {\em does not} reach a certain threshold, or absorbing boundary, within a given time interval \cite{aalen2004survival,Giorgini2020,kearney2021statistics, kearney2021note}. Survival probability can be used to measure the reliability, risk, or extinction of a system \cite{Tsumura2020}.


Survival analysis is a key statistical mechanical tool that is frequently used to study rare events across a broad spectrum of natural and engineering systems. A distinctive characteristic of many of these systems is the presence of periodic forcing \cite{moon2021analytical}, of particular relevance in climate science \cite[e.g.,][]{Ghil2020}. When employing survival analysis in such contexts, it is essential to incorporate the periodicity to ensure accurate interpretation and prediction of the survival probabilities.


Consider several concrete examples from engineering.  Effective dam management requires a comprehensive understanding of the frequency of extreme flooding events. Given the intrinsic, typically seasonal, periodicity of many natural water systems, one must treat these seasonal variations when assessing the probability of extreme events \cite{salas2005correlations}. In the design and maintenance of bridges, it is critical to apply survival analysis techniques that consider the cyclic loads and environmental stressors that bridge superstructures routinely endure. This approach is vital for accurately predicting their lifespan and ensuring structural integrity \cite{nabizadeh2018survival}.



Not only is deriving a reliable expression for survival probability important for predictions, but it is insufficient to merely recognize the existence of periodic forcing.  Rather, it must be intricately woven into the fabric of the statistical model. Therefore, by ensuring that survival probability formulations include the inherent rhythms of the system under investigation, confident predictions will lead to more informed decision-making and resilient system design.


We can obtain the survival probability for such an OU process by solving the equivalent Kolmogorov Backward Equation with an absorbing boundary.  This is the partial differential equation that governs the probability density function (PDF) of the process evolving backward in time. However, finding an exact solution to this equation involves the complex algebra of special functions, making its applicability questionable \cite{ricciardi1988first}. Therefore, rigorous, but simpler, approximate solutions are desirable.


This paper generalizes the results presented in \cite{Giorgini2020} by extending them to {\em periodic non-autonomous} OU processes. The main idea is to divide the entire domain into two regions: an interior region centered on the origin, and a region near the absorbing boundary.  We construct leading-order solutions in each region, which we match using asymptotic methods and then construct a uniformly valid composite solution over the entire domain.  This approach has been used successfully for the more complex potentials found in stochastic resonance \cite{Moon2020,moon2021analytical}.  Moreover, in order to assess the limitations of the asymptotic approach, we introduce a different method that uses an integral relationship between the probability density function of the principal variable and the mean first passage time (also called the first hitting time). The veracity of the two methods is then tested using numerical methods.


\section{First passage problem: Non-autonomous periodic Ornstein-Uhlenbeck processes}\label{sec2}

The first passage time is the time required for a stochastic process $X(t)$ to reach a defined threshold $\beta$, in the {\em first instance}, starting from a given initial value $X(t=0)\equiv x$. The first passage probability distribution function is defined as
\begin{equation}
\zeta(x,t) = \frac{\partial}{\partial t}\text{Prob}(t\le T),
\end{equation}
where $T$ is a random variable denoting the first time at which the system reaches the boundary viz.,
\begin{equation}
T \equiv \textrm{inf}\{t\,:\,X(t)>\beta \mid X(t=0)=x\}.
\end{equation}
The survival probability is the time integral of the first passage time probability distribution.


We study a one-dimensional periodic non-autonomous OU process represented by the following Langevin equation
\begin{equation}
\frac{d{X}(t)}{dt}=-a(t)X(t)+f(t)+\sqrt{2 b(t)}\xi(t),
\label{lang}
\end{equation}
where $\xi(t)$ is Gaussian white noise correlated as $\langle \xi(s) \xi(t)\rangle=\delta(t-s)$, and $a(t),\,b(t)$ and $f(t)$ are periodic functions with periods $T_a,\,T_b$ and $T_f$ respectively. Additionally, $\bar{a} = \frac{1}{T_a}\int_{0}^{T_a}a(s)\,ds\,>0$, and $b(t)>0 \,\forall \,t>0$.

The Fokker-Planck equation corresponding to Eq. \eqref{lang} is
\begin{align}\label{FP}
    \frac{\partial \rho(y,t\mid x,s)}{\partial t}=\frac{\partial}{\partial y}\Big\{[(a(t)y(t)-f(t)]\rho(y,t\mid x,s)\Big\}
    +b(t)\frac{\partial^2\rho(y,t\mid x,s)}{\partial y^2},
\end{align}
with $\rho(y=-\infty,t\mid x,s) = \rho(y=\beta,t\mid x,s)=0$. The boundary
condition $\rho(y=\beta, t\mid x,s)=0$ implies that particles 
passing through $y=\beta$ vanish, and thus slowly disappear over time. For parsimony of notation, in what follows we drop the dependency of $\rho$ on the initial position $x$ and time $s$. 
Moreover, by rescaling Eq. (\ref{lang}) as follows, 
\begin{equation}
X(t) \to X(t)\sqrt{\frac{\bar{b}}{\bar{a}}},\;\;\; f(t) \to \frac{f(t)}{\sqrt{\bar{b}\bar{a}}},\;\;\text{and}\;\; t \to \frac{t}{\bar{a}},  
\end{equation}
we can set $\bar{a}=1$ and $\bar{b}=1$ in Eq. (\ref{FP}). 

\textcolor{black}{Although we require that $b(t)>0$, there are otherwise no specific constraints on the size of the fluctuations of the time periodic coefficients \(a(t)\),  \(b(t)\) and  \(f(t)\) around their mean values. However, we require that \(T \gg 1\), which is equivalent to the probability of the system reaching the boundary remaining small. This condition implies that the fluctuations in the coefficients should not significantly increase the likelihood of boundary crossing.}

Exact analytical solutions of Eq. (\ref{FP}) are unavailable due to the non-autonomous structure of this Fokker-Planck equation.  
Therefore, solving the first passage time problem begins by constructing
an approximate, but asymptotically valid, analytical solution to Eq. (\ref{FP}).
With this in hand, we then calculate the rate at which particles escape at the boundary $X=\beta$.




\section{Analytical methods of calculating the escape rate function of a non-autonomous Ornstein-Uhlenbeck process}

In the next two sections, we describe two different analytical methods of determining the escape rate function of a non-autonomous OU process.  These distinct approaches rely on approximate methods from different disciplines--applied mathematics and statistical physics respectively.  We then compare their accuracy using numerical methods finding the first method to be superior. 

\subsection{Method of Matched Asymptotic Expansions}\label{sec3}

In the method of matched asymptotic expansions one divides a domain into subregions of particular relevance to the problem at hand.  In each region, the governing equation(s) is (are) rescaled according to the dominant processes therein, and an approximate perturbative solution is obtained.  Finally, a uniformly valid composite solution over the entire domain is constructed using asymptotic matching of the solutions in the subregions \cite{Moon2020,moon2021analytical,BenderOrszag}.  Next, we detail this procedure, but refer the reader to the book of Bender and Orszag \cite{BenderOrszag} for a systematic treatment of a wide range of examples.

We begin with the solution of Eq. (\ref{FP}) in the limit $t \gg 1$ with boundary conditions $\rho(\pm\infty,t) =0$, which is
\begin{equation}
\rho_S(y,t)=\frac{\text{e}^{-\frac{[y-F(t)]^2}{2\sigma^2(t)}}}{\sqrt{2\pi\sigma^2(t)}},
\label{rhoS}
\end{equation}
with
\begin{equation}
\sigma^2(t,s)=2\text{e}^{-2\int_{s}^{t}a(r)dr}\int_{s}^{t}b(r)\text{e}^{2\int_{s}^{r}a(u)ds}du,
\label{sigma2}
\end{equation}
and 
\begin{equation}
F(t,s)=\text{e}^{-\int_{s}^{t}a(r)dr}\int_{s}^{t}f(r)\text{e}^{\int_{s}^{r}a(u)du}dr.
\end{equation}
In Eq. (\ref{rhoS}), we write $\sigma^2(t)\equiv\sigma^2(t,0)$ and $F(t)\equiv F(t,0)$.
The solution can be constructed by either using a Fourier transform
with respect to $y$, or by simply using a Gaussian form.  

We let $\rho(y,t)=\rho_S(y,t)\phi(y,t)$, where $\phi(y,t)$ satisfies
\begin{align}
\partial_t \phi(y,t)=-[c(t)y+g(t)] \partial_y\phi(y,t)+b(t)\partial_{yy}\phi(y,t),
\label{Qeq}
\end{align}
in which
\begin{equation}\begin{split}
c(t)&=2\frac{b(t)}{\sigma^2(t)}-a(t),\qquad \text{and}\\
g(t)&=f(t)-2\frac{b(t)}{\sigma^2(t)}F(t).
\end{split}\end{equation} 
For constant coefficients, in the limit $t \to \infty$, we recover $c=a$ and $g=-f$.

\begin{figure}
	\centering
	\includegraphics[width=0.8\textwidth]{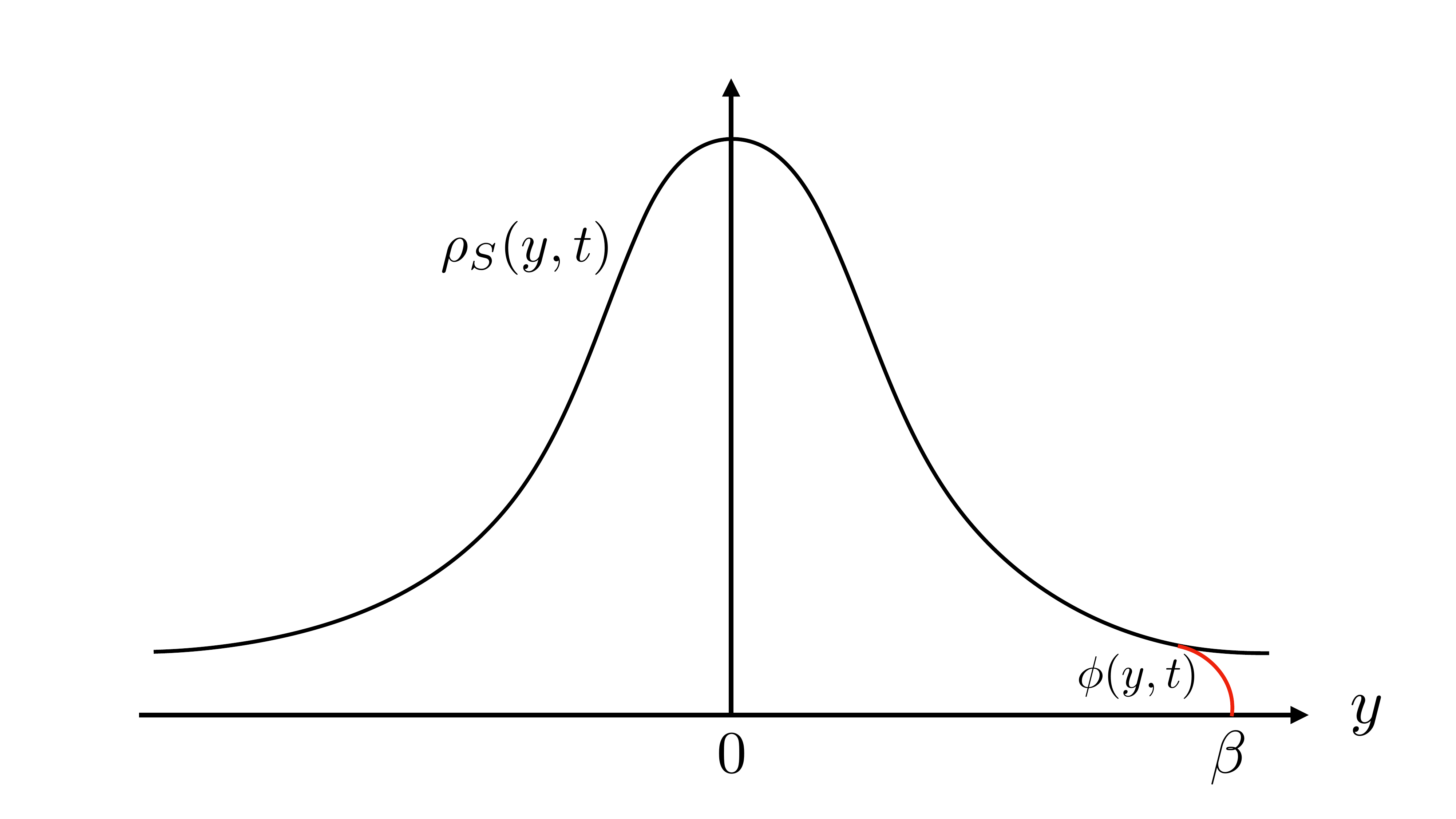}
	\caption{Factorization of $\rho(y,t)$ with an absorbing boundary at $y=\beta$. The black curve is $\rho_S(y,t)$ while the red curve is $\phi(y,t)$.}
	\label{Fig1}
\end{figure} 

\begin{figure}
	\centering
	\includegraphics[width=0.8\textwidth]{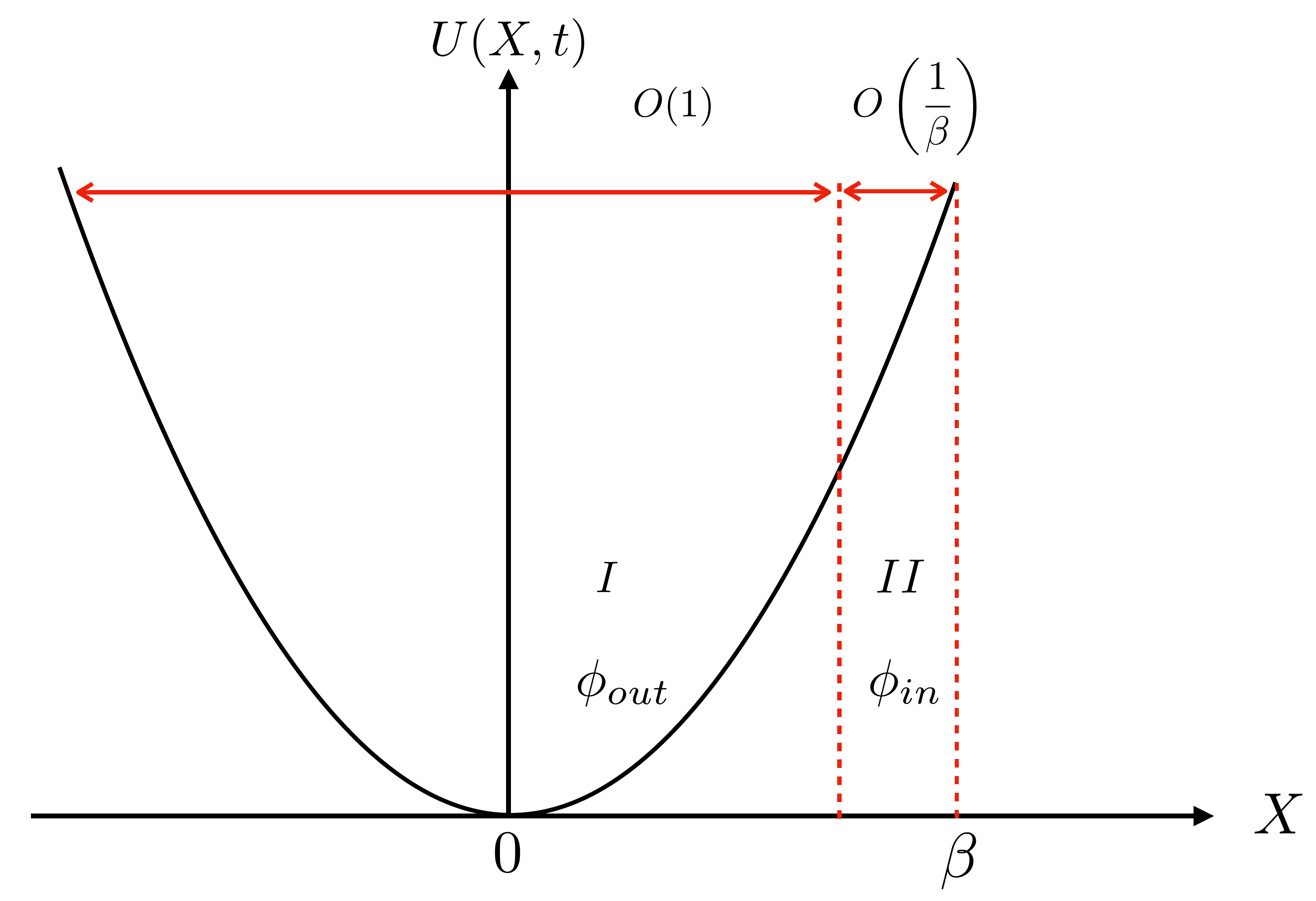}
	\caption{Schematic of the non-autonomous Ornstein-Uhlenbeck problem addressed. We divide the time-dependent potential $U(X,t) = \frac{1}{2} a(t) X^2-f(t) X$
        into two regions: a broad $O(1)$ region ($I$) that contains the minimum of the potential $X = 0$, and a narrow $O(1/\beta)$ boundary layer region ($II$) near $X = \beta$, where $\beta \gg 1$.  In each region we find the asymptotically dominant solutions ($\phi^{out}$ in $I$ and $\phi^{in}$ in $II$) and then match them as described in Section \ref{sec3}.}
	\label{Fig2}
\end{figure}

Because the probability of a Brownian particle reaching the boundary at $X=\beta$ in such an OU process is quite small, it is reasonable to assume that $\beta \gg 1$ to capture the fact that the event 
is rare.  Therefore, we divide the domain into two regions: a broad $O(1)$ region ($I$) that contains the minimum of the potential, $X = 0$, and a narrow $O (1/\beta )$ boundary layer ($II$) near $X = \beta$, as shown
in Fig. \ref{Fig2}.
We solve the limiting differential equations within these two regions, from which we construct an approximate uniform solution by asymptotic matching. We denote $\phi_{out}$ and $\phi_{in}$ as the solutions in regions $I$ and $II$, respectively.  Importantly, we emphasize that in the parlance of matched asymptotic analysis, the solutions within the boundary layer are the ``inner solutions'' and those outside the boundary layer are the ``outer solutions''.  They {\em do not} refer to the inner and outer parts of the potential itself.  In fact, more generally, a boundary layer can appear anywhere in the domain in which one seeks solutions. 

In region $I$, the outer solution $\rho_{out}$ satisfies
\begin{align}
 \frac{\partial \rho_{out}(y,t)}{\partial t}=\frac{\partial}{\partial y}\Big\{[(a(t)y(t)-f(t)]\rho_{out}(y,t)\Big\}
    +b(t)\frac{\partial^2\rho_{out}(y,t)}{\partial y^2},
\end{align}
with the boundary conditions $\rho_{out}(y=-\infty,t)=0$ and $\rho_{out}(y=\infty,t)=0$. 
The outer solution does not satisfy the boundary condition at $y=\beta$, and is
proportional to $\rho_S(y,t)$. Hence, we let $\rho_{out}(y,t)=\rho_S(y,t)\phi_{out}(t)$, 
where $\phi_{out}(t)$ represents the slow decrease in probability from the leakage to the boundary $y=\beta$.  Therefore, we must seek a solution that vanishes on the boundary for all time, and outside the boundary layer approaches $\phi_{out}$ asymptotically. 


In the boundary layer (region $II$), where the dominant balances in the governing equation change, the solution decreases abruptly to zero to satisfy the boundary condition. 
In particular, the diffusive term, associated with the second derivative with respect to $y$, will
be dominant in the boundary layer where the gradients are the steepest.  Now we develop this in detail.


Because the gradients are steep in a small region $y=O(1/\beta)$, we let $\epsilon\equiv 1/\beta \ll 1$, and introduce a stretched coordinate as $\eta=(y-1/\epsilon)/\epsilon$. Thus, in the inner region $\phi_{in}=\phi_{in}(\eta,t)$, which is governed by the following form of Eq. (\ref{Qeq}); 
\begin{align} 
\label{eq:qbd}
\epsilon^2[\partial_t\phi_{in}(\eta,t)
+c(t)\eta\partial_{\eta}\phi_{in}(\eta,t)]&+ 
[c(t)+\epsilon g(t)]\partial_\eta \phi_{in}(\eta,t)\nonumber \\
&=b(t)\partial_{\eta\eta}\phi_{in}(\eta,t),
\end{align}
with $\phi_{in}(\eta=0,t)=0$ and $\phi_{in}(\eta=-\infty,t)= K <  \infty$, and $K$ is a constant to be determined as part of the procedure to asymptotically match $\phi_{in}$ to the outer solution $\phi_{out}$. Ignoring $O(\epsilon^2)$ terms, Eq. \eqref{eq:qbd} becomes
\begin{align}
 [c(t)+\epsilon g(t)]d_\eta \phi_{in}(\eta,t)=b(t)d_{\eta\eta}\phi_{in}(\eta,t), 
 \label{dom_balance}
\end{align}
the solution of which is
\begin{equation}\begin{split}
\phi_{in}(\eta,t)&=K\left(1-\text{exp}\left[\left(\frac{c(t)+\epsilon g(t)}{b(t)}\right)\eta\right] \right) \\
&\equiv K \phi^\prime_{in}(\eta,t)
\label{leadingQ}
\end{split}
\end{equation}
The solution of Eq. \eqref{leadingQ} is valid when
$\lim_{\eta\to -\infty}\phi_{in}$ is bounded, for which we require $c(t)+\epsilon g(t) > 0$ for all $t$.  We focus on this situation in the remainder of this section.  
However, as discussed in detail in Appendix \ref{secA2}, for
a specific time $t^*$ such that $c(t^*)+\epsilon g(t^*)<0$,
this solution inside the boundary layer is no longer valid (cf. Eq. \ref{AppleadingQ}).

Asymptotic matching requires that the outer limit of the inner solution equal the inner limit of the outer solution, but the outer solution depends only on time, so that 
\begin{equation}
\lim_{\eta\to-\infty}\phi_{in}(\eta,t) = \phi_{out}(t), \\
\qquad \text{giving}\qquad K=\phi_{out}(t).
\label{eq:matching}
\end{equation}
The uniformly valid composite solution is the sum of the inner and the outer solutions minus the common part viz.
\begin{align}
 \phi(y,t) &=\phi_{out}(t)+\phi_{in}(y,t)-\lim_{\eta\to-\infty}\phi_{in}(\eta,t) \nonumber  \\
&=\phi_{out}(t)\phi^\prime_{in}(y,t).
\end{align}
This implies that
\begin{equation}\begin{split}
\rho(y,t)&=\rho_S(y,t)\phi(y,t)=\rho_S(y,t)\phi_{out}(t)\phi^\prime_{in}(y,t)\\
& = \phi_{out}(t)\phi^\prime_{in}(y,t) \frac{1}{\sqrt{2\pi\sigma^2(t)}}\text{exp}\left(-~\frac{[y-F(t)]^2}{2\sigma^2(t)}\right). 
\label{eq:rho_matching}
\end{split}
\end{equation}
Now, integrating the Fokker-Planck equation (\ref{FP}) gives
\begin{equation}
\partial_t\int_{-\infty}^{\beta}\rho(y,t)dy\simeq \partial_t\left(\phi_{out}(t)\int_{-\infty}^{\infty}\rho_S(y,t)dy\right)=b(t)\partial_y \rho(y,t)\mid_{y=\beta},
\end{equation}
which leads to the time-evolution of $\phi_{out}$ as $d\phi_{out}/dt\equiv-r(t)\phi_{out}$, where the escape rate function is
\begin{align}
r(t)= \frac{\beta c(t)+g(t)}{\sqrt{2\pi\sigma^2(t)}}\text{exp}\left(-~\frac{[\beta-F(t)]^2}{2\sigma^2(t)}\right),\;\;\; \forall ~~ \beta c(t)+g(t)~ > ~0.
\label{rate1}
\end{align}


\subsection{Integral method}\label{sec4}

In this section, we will present a different approach to estimating the survival probability of the non-autonomous OU process with time-periodic coefficients given by Eq. (\ref{lang}).

The probability density that the process starting at $X(t=0)=0$ reaches $\beta$ at time $t$ can be written as
\begin{equation}
\rho(\beta,t\mid 0,0)=\int_0^t dr \, \zeta(\beta,r\mid 0,0)\rho(\beta,t\mid\beta,r),
\label{rho_ren}
\end{equation}
where $\zeta$ is the first passage time PDF, and
\begin{equation}
\rho(\beta,t\mid \beta,s)=\frac{\text{e}^{-\beta^2\frac{\left(1-m(t,s)-\frac{F(t,s)}{\beta}\right)^2}{2\sigma^2(t,s)}}}{\sqrt{2\pi\sigma^2(t,s)}},
\end{equation}
where $\sigma^2(t,s)$ is defined in Eq. (\ref{sigma2}) and we let
\begin{equation}
m(t,s)= \text{e}^{-\int_{s}^{t} a(r)\,dr}.
\end{equation}
Thus, we write the probability density in Eq.~\eqref{rho_ren} as
\begin{equation}\begin{split}
\rho(\beta,t\mid0,0)=&\int_0^t dr\dfrac{\text{e}^{-\beta^2 S(t,r)}}{\sqrt{2\pi\sigma^2(t,r)}} \zeta(\beta,r\mid0,0)\\
=&\int_{t-\delta}^t dr\dfrac{\text{e}^{-\beta^2 S(t,r)}}{\sqrt{2\pi\sigma^2(t,r)}} \zeta(\beta,r\mid0,0)\\
&+\int_0^{t-\delta} dr\dfrac{\text{e}^{-\beta^2 S(t,r)}}{\sqrt{2\pi\sigma^2(t,r)}} \zeta(\beta,r\mid0,0)\\
=&\int_{t-\delta}^t dr\dfrac{\text{e}^{-\beta^2 S(t,r)}}{\sqrt{2\pi\sigma^2(t,r)}} \zeta(\beta,r\mid0,0)\\
&+\rho_S(\beta,t)\int_0^{t-\delta} dr \zeta(\beta,r\mid0,0),
\end{split}\end{equation}
for a $\delta$ such that $1\ll \delta \ll t$, where $\rho_S$ is given in Eq. (\ref{rhoS}), and 
\begin{equation}
S(t,s)\equiv\dfrac{\left(1-m(t,s)-\frac{F(t,s)}{\beta}\right)^2}{2\sigma^2(t,s)}.
\end{equation}

Now, consider the first integral in $\rho(\beta,t\mid0,0)$. Because $\delta \ll t$ and $\zeta$ is slowly varying with time, the $\zeta$ term can be taken outside of the integral.  We then expand both $S(t,s)$ and $\sigma^2(t,s)$ about $s\approx t$ to find the following at leading order in $t-s$;
\begin{equation}\begin{split}
\lim_{s \to t} S(t,s) &= \frac{\left[a(t)-\frac{f(t)}{\beta}\right]^2}{4 b(t)}(t-s), \qquad \text{and}\\
\lim_{s \to t} \sigma^2(t,s) &= 2b(t)(t-s),
\end{split}\end{equation}
where we used $\lim_{s \to t} \frac{F(t-s)}{t-s}=f(t)$.
Taking $\beta \to \infty$, and substituting these expressions in the first integral, we obtain
\begin{equation}\begin{split}
\int_{t-\delta}^t dr\dfrac{\text{e}^{-\beta^2 S(t,r)}}{\sqrt{2\pi\sigma^2(t,r)}} \zeta(\beta,r\mid0,0)&= \zeta(\beta,t\mid0,0)\int_{-\delta}^0 dr\dfrac{\text{e}^{\beta^2 \frac{\left(a(t)-\frac{f(t)}{\beta}\right)^2}{4 b(t)}r}}{\sqrt{-4\pi b(t)r}}\\
&= \frac{\zeta(\beta,t\mid0,0)}{\beta \left(a(t)-\frac{f(t)}{\beta}\right)}\textrm{erf}\bigg[\frac{\beta\sqrt{\delta}\left(a(t)-\frac{f(t)}{\beta}\right)}{2\sqrt{b(t)}}\bigg]\\&\simeq \frac{\zeta(\beta,t\mid0,0)}{\beta \mid a(t)-\frac{f(t)}{\beta}\mid}.
\end{split}\end{equation}

Now, for $t \gg 1$, we can write Eq.~(\ref{rho_ren}) as
\begin{equation}
\frac{\rho(\beta,t\mid0,0)}{\rho_S(\beta,t)}\simeq 1\simeq \frac{\zeta(\beta,t\mid0,0)}{r(t)}+ \int_{0}^{t}dr\zeta(\beta,r\mid0,0), 
\end{equation}
where $r(t)=\beta \mid a(t)-\frac{f(t)}{\beta}\mid\rho_S(\beta,t)$.
Taking the partial derivative with respect to time on both sides we find
\begin{equation}\begin{split}
& 0\simeq \frac{\partial_t \zeta(\beta,t\mid0,0)}{r(t)}+\zeta(\beta,t\mid0,0)\left[\partial_t \frac{1}{r(t)}+ 1\right], \text{~~so that}\\
\partial_t \zeta(\beta,t\mid0,0)&\simeq -\zeta(\beta,t\mid0,0) \left[\partial_t \ln \frac{1}{r(t)}+r(t)\right], \text{~~and hence} \\
\zeta(\beta,t\mid0,0)& \simeq \text{e}^{-\int_0^tdu\left[\partial_u \ln \frac{1}{r(u)}+ r(u)\right]}\\ 
&= \text{e}^{-\int_0^t du r(u)}r(t).
\label{surv_na}
\end{split}\end{equation}
Finally, we can identify $r(t)$ with the rate function
\begin{equation}
r(t) =  \frac{\mid a(t)-\frac{f(t)}{\beta}\mid\beta}{\sqrt{2\pi\sigma^2(t)}}\text{e}^{-\frac{(\beta-F(t))^2}{2\sigma^2(t)}}.
\label{rate2}
\end{equation}
Here, however, we have no constraints on the sign of the coefficients.

When the model coefficients \(a\), \(b\), and \(f\) are constant, the expressions for \(F(t)\) and \(\sigma^2(t)\) are
\begin{equation}
F(t) = \frac{f}{a}\left(1-\text{e}^{-a t}\right), \text{~~and~~}
\sigma^2(t) = \frac{b}{a}\left(1-\text{e}^{-2 a t}\right),
\end{equation}
so that for \(t \gg 1\), the escape rate function simplifies to
\begin{equation}
r(t) =  \sqrt{\frac{a}{2\pi b}}\left(\beta a-f\right)\text{e}^{-\frac{a\left(\beta-\frac{f}{a}\right)^2}{2b}t}.
\label{rateA}
\end{equation}
Therefore, we recover the well-established expression for the escape rate function of an {\em autonomous} Ornstein-Uhlenbeck process (see e.g., \cite{Giorgini2020}).

\section{Comparing the two methods}\label{sec5}

In Sections \ref{sec3} and \ref{sec4} we used two distinct methods to derive expressions for the escape rate function of a non-autonomous OU process.  In this section we compare them.  

We begin by writing the escape rate function as 
\begin{equation}
r(t) =  \frac{\mid h_{1,2}(t)\mid\beta}{\sqrt{2\pi\sigma^2(t)}}\text{e}^{-\frac{(y-F(t))^2}{2\sigma^2(t)}},
\label{rate12}
\end{equation}
where $h_1(t)=c(t)+\frac{g(t)}{\beta}$ for the first method, and $h_2(t)=a(t)-\frac{f(t)}{\beta}$ for the second method. These expressions are identical for constant coefficients and $t \gg 1$.


To compare the two methods, we solved the Kolmogorov Backward Equation for the survival probability $S(x,t)$ of this non-autonomous OU process numerically, which is
\begin{equation}
\partial_t S(x,t) = -[a(t)x-f(t)]\partial_xS(x,t)+b(t)\partial_{xx}S(x,t).  
\label{surv_eq}
\end{equation}
The initial condition is $S(x,0)=\Theta(\beta-x)$, where $\Theta(\cdot)$ is the Heaviside theta function, 
with boundary conditions $S(-\infty,t)=1,\, S(\beta,t)=0\, ~\forall~ t>0$ (see \cite{Giorgini2020} for more detail). For $\beta \gg 1$ we approximate $S(0,t)$ as
\begin{equation}
S(0,t)\approx \text{e}^{-\int_0^t r(s)ds},
\label{S0}
\end{equation}
and compute the root mean squared error between (a) the numerical estimate of the survival probability $S_{\textrm{num}}(0,t)$, for all time less than a maximum value, $t_{max}$, chosen such that $S_{\textrm{num}}(0,t)~<~0.1~ \, \forall ~t~<~t_{max}$, and (b) the approximate analytical expressions $S_{\textrm{an}}(x,t)$, which use the rate functions obtained by the first ($\textrm{RMSE}_1$), and second method ($\textrm{RMSE}_2$). We generated $N=100$ different processes with time-periodic coefficients of the form
\begin{equation}
\alpha(t) = \alpha_0 + \alpha_1 \\\sin(\omega_\alpha t) + \alpha_2 \\\cos(\omega_\alpha t),
\end{equation}
where $\alpha_{1,2}$ have been randomly chosen in the interval $[-1,1]$, subject to the constraint $\sqrt{\alpha_1^2+\alpha_2^2}=\alpha_0 K$, with $K\in [0,1]$. For both $a(t)$ and $b(t)$ we set $\alpha_0=1$ since their time averages are $\bar{a}=\bar{b}=1$. We choose $f(t)$ randomly in the interval $[0,1]$, and the frequency $\omega_\alpha$ is randomly chosen such that $\omega_\alpha>2\pi/50$. 
 
\begin{figure}
	\centering
	\includegraphics[width=1.\textwidth]{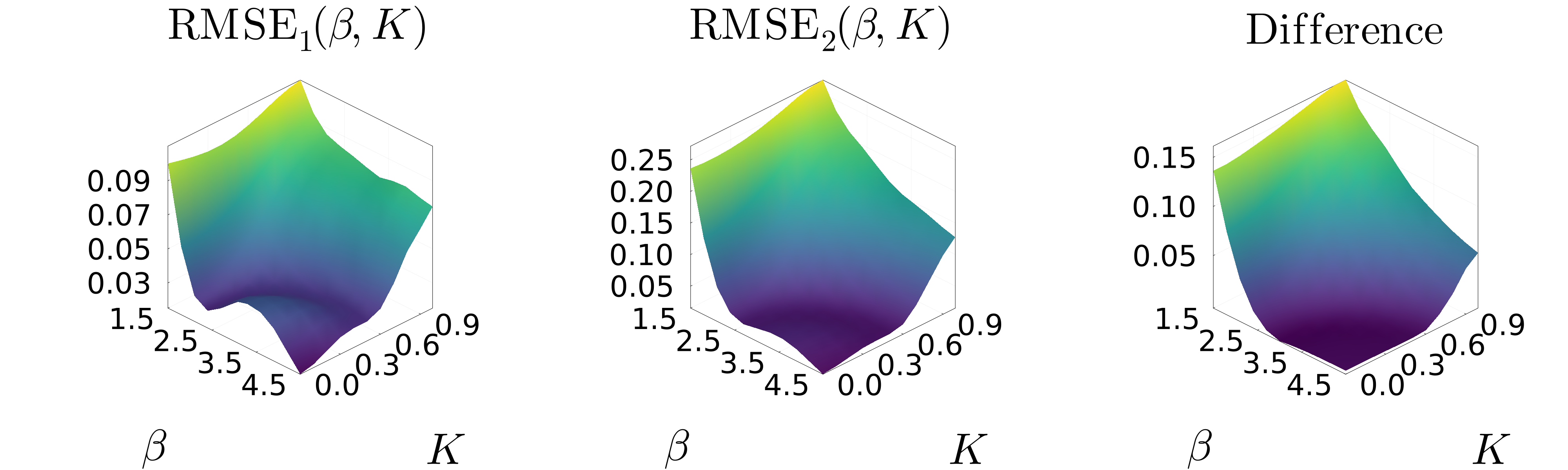}
	\caption{The root mean square error between the numerical ($S_{\textrm{num}}(0,t)$) and the analytical ($S_{\textrm{an}}(0,t)$) solutions for the survival probability for different values of $\beta$ and $K$. From left to right: $S_{\textrm{an}}(0,t)$ is obtained from the first method ($\textrm{RMSE}_1$), from Section \ref{sec3}, the second method ($\textrm{RMSE}_2$), from Section \ref{sec4}, and their difference (Difference).}
\label{Fig3}
\end{figure} 

Figure~(\ref{Fig3}) shows the root mean square error between the numerical and both analytical solutions, and their difference, for the survival probability as a function of the parameters $\beta$ and $K$.  The accuracy of the analytical estimates improves as $\beta$ increases, because it reflects the limit $\beta\to \infty$ that we used to construct the solutions. On the other hand, increasing the weight of the time-dependent contribution, represented by the parameter $K$, increases the error, reflecting the challenge of capturing the non-autonomous features of the system. Clearly, the method of matched asymptotic expansions from Section \ref{sec3} is superior, particularly in the parameter regime where the numerical and analytical solutions agree less well; for large values of $K$ and small values of $\beta$.


\section{Analytical expression for the survival probability}\label{sec6}

Given the results of Section \ref{sec5}, we now apply the asymptotic method presented in Section \ref{sec3} to obtain an analytical solution of Eq.~(\ref{surv_eq}) over the entire $x$-domain. We note here again, that $\beta \gg 1$ and we divide the domain into two regions: a broad $O(1)$ region ($I$) that contains the minimum of the potential, $X = 0$, and a narrow $O(1/\beta)$ boundary layer region ($II$) near $X = \beta$, as shown in Fig.~\ref{Fig2}. We solve the limiting differential equations within the two regions, from which we construct an approximate composite uniform solution by asymptotic matching. 
The solutions in regions $I$ and $II$ are $S_{out}$ and $S_{in}$ respectively. The outer solution $S_{out}$ is given by Eq.~(\ref{S0}), and using the same procedure that led to Eq.~(\ref{leadingQ}), we arrive at the composite uniformly continuous asymptotic solution as
\begin{equation}
S(x,t) = \text{e}^{-\int_0^t r(s)ds}\left(1-\text{exp}\left[\frac{\beta a(t)-f(t)}{b(t)}\beta(x-\beta) \right]\right),
\label{eq_S1}  
\end{equation}
where we have considered times such that $t\gg 1/\beta^2$.  Recall that $T_a,\,T_b$ and $T_f$ are the periods of 
$a(t),\,b(t)$ and $f(t)$ respectively.  Crucially, then, if $T_a, T_b, T_f \gg 1/\beta^2$, then for time scales of order $1/\beta^2$, the survival probability can be approximated by the autonomous result $S_{out}(x,t)$ of Giorgini et al. \cite{Giorgini2020}. 

Finally, we arrive at an expression for the survival probability that reduces to $S_{out}(x,t)$ for $t \to 0^+$ and to Eq.~(\ref{eq_S1}) for $t \gg 1/\beta^2$, which is       
\begin{align}
&S(x,t)=\text{e}^{-r(t)}\Bigg(\frac{1}{2}\text{erfc}
\left[\frac{\beta\left(x-\beta\right)-\left(a(0)-\frac{f(0)}{\beta}\right)t\beta^2}{\sqrt{4b(0)t\beta^2}}\right]  \nonumber \\
&+\frac{1}{2}\text{exp}\left[\frac{\left(a(t)-\frac{f(t)}{\beta}\right)\beta^2}{b(t)\beta^2}\beta(x-\beta)\right]
\text{erfc}\left[-\frac{\beta\left(x-\beta\right)+\left(a(0)-\frac{f(0)}{\beta}\right)t\beta^2}{\sqrt{4b(0)t\beta^2}}\right]\Bigg), 
\label{Surv_short}
\end{align}
where $\text{erfc}(x)$ is the complimentary error function. 

\begin{figure}
	\centering
	\includegraphics[width=1.\textwidth]{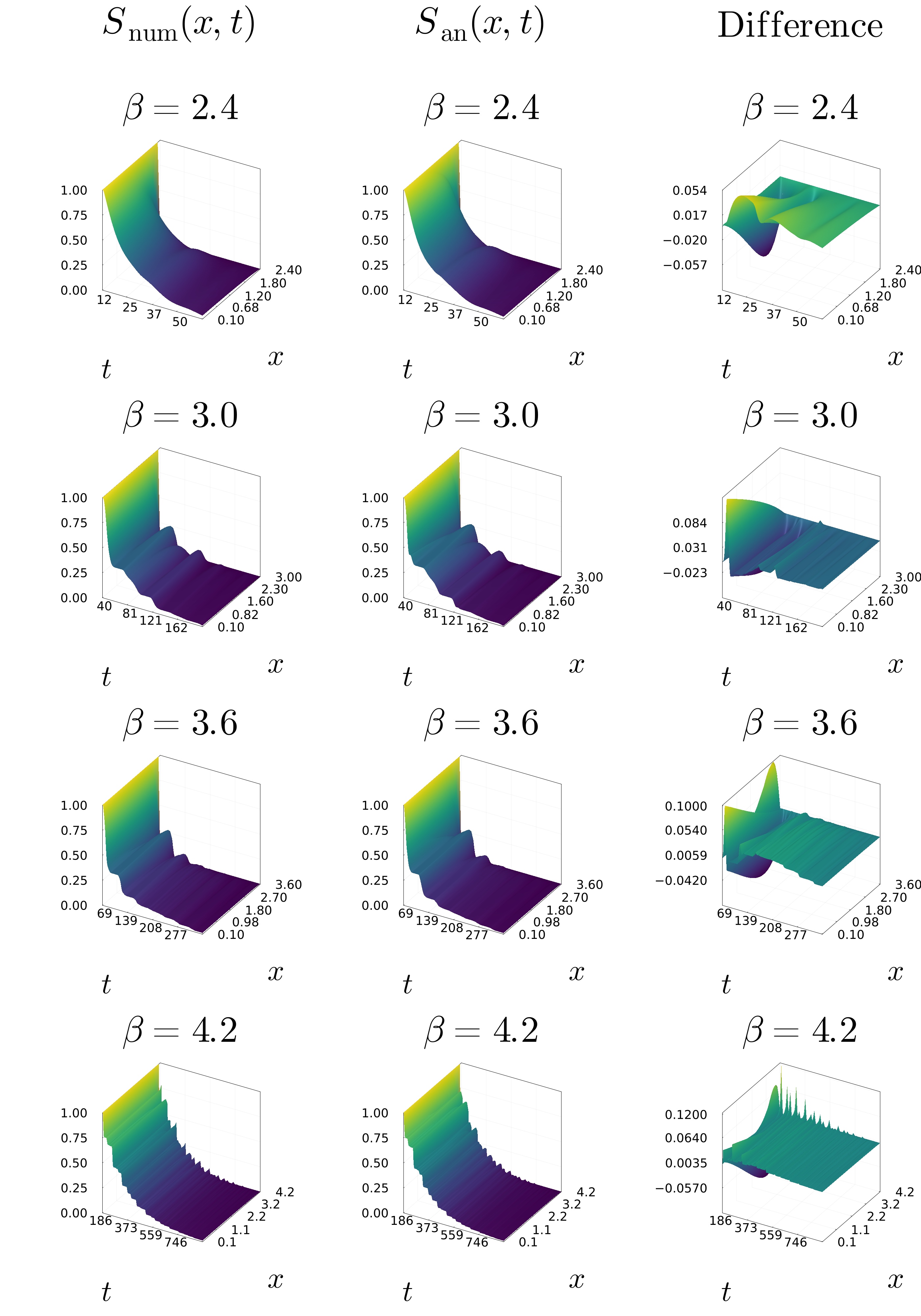}
	\caption{Plot of $S_{\textrm{num}}(x,t)$, $S_{\textrm{an}}(x,t)$ and their difference in function of $x$ and $t$. In each row, we used a different value of $\beta$ and different coefficients as reported in Appendix \ref{secA3}.}
\label{Fig4}
\end{figure}

In Fig. (\ref{Fig4}) we compare the analytical (Eq.~\ref{Surv_short}) and numerical results for the survival probability, $S_{\textrm{an}}(x,t)$ and $S_{\textrm{num}}(x,t)$ respectively, as well as their difference, as a function of $x \in [0,\beta]$ and $t$, for different values of $\beta$ and the time-dependent coefficients.  As discussed in Section \ref{sec5} the asymptotic method described in Section \ref{sec3} is superior to the method of Section \ref{sec4}.  Therefore, we use $h_1(t)$ in Eq.~(\ref{rate12}), and hence in Eq.~\eqref{Surv_short}, rather than $h_2(t)$, and this is the $S_{\textrm{an}}(x,t)$ plotted. We note that the analytical estimates agree very well with the numerical solution, even when $\beta$ is not very large.  In Appendix \ref{secA1} we treat the Kolmogorov Backward Equation (\ref{surv_eq}) for time scales $t=O(1/\beta^2)$.


\section{Conclusions}\label{sec7}

In this article we derived formulae for the survival probability of a periodic non-autonomous Ornstein-Uhlenbeck (OU) process using two distinct methodologies. The first approach uses asymptotic methods to solve the associated Kolmogorov Backward Equation with an absorbing boundary. The approach (i) divides the domain into an inner region near the boundary, and an outer region away from the boundary; (ii) rescales the governing equation in these two regions and determines the leading order solutions in each, and; (iii) asymptotically matches these solutions to construct a continuous composite solution. The second approach analyzes the integral relationship between the probability density function and the mean first passage time probability density function, which in turn provided an approximate form for the exit rate. Both approximate analytical solutions were tested by comparison with numerical solutions, and computing the root mean squared error (RMSE) between the two.  
We found that the first asymptotic matching approach is superior, particularly in regions of the parameter space where achieving agreement between numerical and analytical solutions proved challenging.
\textcolor{black}{This superiority arises from the nature of the asymptotic method, in which a uniformly valid composite solution that accurately captures the behavior in both the interior region and the boundary layer is constructed. In particular, the method rigorously connects the steep gradients near the absorbing boundary with the remainder of the domain. In contrast, the integral method relies on approximations that do not fully capture the behavior of the solution near the boundary, especially when the coefficients are time-dependent. In consequence, the asymptotic method provides a more accurate approximation, particularly in regimes where boundary layer effects are significant.}



Clearly, there are many advantages of circumventing the complexities associated with finding an exact solution of the Kolmogorov Backward Equation.  Moreover, a simple, approximate, and yet highly accurate analytical expression for the survival probability offers a significant advantage over purely numerical approaches, and thus is of use across the wide range of applications in natural and engineering systems where survival analysis is pivotal. For instance, in climate science or in engineering where understanding the frequency of extreme events or component failure are central, our approach can be used both as a predictor or as a framework for data analysis. 

Finally, this study provides a solid foundation for future explorations into other stochastic processes.  In particular, 
it is hoped that having demonstrated the power of this asymptotic method, our approach may be more widely adopted in the challenging area of non-autonomous stochastic systems.  The simplicity and accuracy of the approach make it a robust tool for further research into, and practical applications of, survival analysis.



\bmhead{Acknowledgments}

We gratefully acknowledge support from the Swedish Research Council Grant No. 638-2013-9243.

\section*{Data Availability} 

This manuscript has no associated data beyond that produced by solving the equations herein.

\begin{appendices}

\section{Survival probability when the model coefficients share the same time dependence}\label{secA1}

Here we consider the particular case where all of the time-dependent coefficients appearing in the Langevin Eq. (\ref{lang}) share the same time dependence; $a(t)=a_0 h(t)$, $b(t)=b_0 h(t)$ and $f(t)=f_0 h(t)$, with amplitudes $a_0$, $b_0$ and $f_0$, for a periodic function $h(t)$ with period $T$ such that $\frac{1}{T}\int_0^Tdt~ h(t)>0$. 

Consider the Kolmogorov Backward Equation (\ref{surv_eq}) in the boundary layer region ($II$), and introduce a rescaled time coordinate, $\theta=t/\epsilon^2$, and stretched coordinate $\eta=(x-\frac{1}{\epsilon})\frac{1}{\epsilon}$, where $\epsilon \ll 1$ as in Section (\ref{sec3}). The leading-order ($1/\epsilon^2$) balance of the rescaled inner equation becomes
\begin{align} 
\partial_\theta S_{in}(\eta,\theta)=-[a(\theta)+\epsilon f(\theta)] \partial_\eta S_{in}(\eta,\theta)+b(\theta)\partial_{\eta\eta} S_{in}(\eta,\theta),
\label{eq:eqg0}
\end{align}
with boundary conditions $S_{in}(\eta=0,\theta)=0$ and $S_{in}(\eta=-\infty,\theta)=K_1<\infty$, and initial condition $S_{in}(\eta,\theta=0)=\Theta(-\eta)$, where again $\Theta(\cdot)$ is the Heaviside theta function. 

We first consider a function $G$ satisfying Eq. (\ref{eq:eqg0}) with the same boundary conditions but different initial condition $G(\eta,0)=\delta(\eta-\eta_0)$. This allows us to introduce characteristics governed by 
\begin{equation}
\frac{d\eta}{d\theta}=a(\theta)+\epsilon f(\theta) \;\;\;\textrm{and}\;\;\; \frac{d\rho}{d\theta} = 1.
\end{equation}
Let $\mu=\eta-\int_{0}^{\theta}[a(s)+\epsilon f(s)]ds\equiv\eta-\alpha(\theta)$, and then Eq. (\ref{eq:eqg0}) for $G$ becomes
\begin{align}   
\partial_\theta G(\mu,\rho)=b(\theta)\partial_{\mu\mu}G(\mu,\rho),
\label{eq:eqg1}
\end{align}
where one boundary condition becomes $G(\mu=-\alpha(\theta),\theta)=0$.

A general solution of Eq. ({\ref{eq:eqg1}}) is
\begin{equation}\begin{split}
\Psi(\mu,\theta)=\frac{1}{\sqrt{4\pi\gamma(\theta)}}
\text{exp}\left[-\frac{(\mu-k_1)^2}{4\gamma(\theta)}+k_2\right], 
\label{eq:eqg2}
\end{split}\end{equation}
where  $\gamma(\theta)=\int_{0}^{\theta}b(s)ds$.  In order to satisfy Eq. (\ref{eq:eqg1}), we must have $k_1$ and $k_2$ be constants with respect to $\theta$ and $\mu$.


A particular solution of Eq. ({\ref{eq:eqg1}}) that satisfies the initial and boundary conditions in the new coordinates is a linear combination viz.,
\begin{equation}\begin{split}
&G(\mu,\theta;\mu_0)=\Psi_1(\mu,\theta;\mu_0)+\Psi_2(\mu,\theta;\mu_0), \\
&\Psi_1(\alpha(\theta),\theta;\mu_0)+\Psi_2(\alpha(\theta),\theta;\mu_0)=0, 
\label{eq:eqg23}
\end{split}\end{equation}
with the right choice of $k_1$ and $k_2$, where
\begin{align}
&\Psi_1(\mu,\theta;\mu_0)= \frac{1}{\sqrt{4\pi\gamma(\theta)}}
\text{exp}\left[-\frac{(\mu_0-\mu)^2}{4\gamma(\theta)}\right], \nonumber \text{~~~and}\\
&\Psi_2(\mu,\theta;\mu_0)= -\frac{1}{\sqrt{4\pi\gamma(\theta)}}
\text{exp}\left[-\frac{(\mu+\mu_0)^2}{4\gamma(\theta)}-\mu_0\frac{\alpha(\theta)}{\gamma(\theta)}\right].
\label{eq:eqg3}
\end{align}


Because $a(t)$, $b(t)$ and $f(t)$ have the same time dependence, then  $\frac{\mu_0\alpha(\theta)}{\gamma(\theta)}=\frac{\mu_0(a_0-\epsilon f_0)}{b_0}$ is constant.  Therefore, Eqs. (\ref{eq:eqg3}) and (\ref{eq:eqg23}) provide the solution to Eq. (\ref{eq:eqg1}).

In order to satisfy the boundary conditions in Eq. (\ref{eq:eqg0}), we define $S_{in}(\eta,\theta)=K_1\int_{0}^{\infty}G(\mu,\theta;\mu_0)d\mu_0$. The explicit form of $S_{in}(\eta,\theta)$ is 
\begin{align}
S_{in}(\eta,\theta)&=K_1 \Bigg[\frac{1}{2}\text{erfc}
\left(\frac{\eta-\alpha(\theta)}{\sqrt{4\gamma(\theta)}}\right)  \nonumber \\
&+\frac{1}{2}\text{exp}\left(\frac{a_0}{b_0}\eta\right)
\text{erfc}\left(-\frac{\eta+\alpha(\theta)}{\sqrt{4\gamma(\theta)}}\right)\Bigg].
\label{eq:eqg5}
\end{align}

The matching condition, $\lim_{\eta\to -\infty}S_{in}(\eta,\theta)=S_{out}(\theta)$, leads to $K_1=S_{out}(\theta)$. Therefore, as in Section \ref{sec3}, the uniformly valid composite solution is the sum of the inner and the outer solutions minus the common part, and is
\begin{align}
&S(x,t)=S_{out}(t)+S_{in}(x,t)-K_1 \nonumber \\
&=\text{e}^{-r(t)}\Bigg[\frac{1}{2}\text{erfc}
\left(\frac{\beta(x-\beta)-\alpha(t\beta^2)}{\sqrt{4\gamma(t\beta^2)}}\right)  \nonumber \\
&+\frac{1}{2}\text{exp}\left(\frac{a_0}{b_0}\beta(x-\beta)\right)
\text{erfc}\left(-\frac{\beta(x-\beta)+\alpha(t\beta^2)}{\sqrt{4\gamma(t\beta^2)}}\right)\Bigg],
\label{eq:eqg51}
\end{align}
expressed using the original variables $x,t$.

\section{Boundary layer method for negative $h_1(t)$}\label{secA2}

Here, we demonstrate how the escape rate function in Eq. (\ref{rate1}) can be adapted to account for the case in which \( h_1(t) = c(t) + \epsilon g(t) \leq 0 \).

We begin with the dominant balance equation in the boundary layer, as given by Eq. (\ref{dom_balance}), which we rewrite here as
\begin{equation}
-\mid c(t) + \epsilon g(t) \mid d_{\eta} \phi_{in}(\eta,t) = b(t) d_{\eta\eta} \phi_{in}(\eta,t).
\label{Appinn}
\end{equation}

Given that in the interior region \( \rho = \rho_S \phi_{in} \), and considering that a probability density function must be non-negative, we impose \( \phi_{in}(\eta,t) \geq 0 \) for all \( t > 0, \eta < 0 \) to ensure that \( \rho > 0 \). Thus, the solution of Eq. \eqref{Appinn} is
\begin{equation}
\phi_{in}(\eta,t) = -K \left( 1 - \exp \left[ -\left(\frac{ \mid c(t) + \epsilon g(t) \mid }{b(t)}\right) \eta \right] \right).
\label{AppleadingQ}
\end{equation}

We denote \( \Delta t_i = t_{i+1} - t_i \) as the \(i\)-th time interval between \( t = t_i \) and \( t = t_{i+1} \) where \( h_1(t) = c(t) + \epsilon g(t) \leq 0 \) holds. 

Eq. (\ref{Qeq}) is the Kolmogorov Backward Equation for a stochastic process with a drift term \( c(t) + \epsilon g(t) \), we consider each interval \( \Delta t_i \) to be sufficiently small to ensure that the event \( X = \beta \) remains rare. 
Owing to Eq. (\ref{eq:matching}),
\begin{equation}
\begin{split}
K(t_i) &= \phi_{out}(t_i), \\
K(t_{i+1}) &= \phi_{out}(t_{i+1}),
\end{split}
\end{equation}
and since $ K $ is a continuous and slowly varying function of $ t $, we expect that for $ t \in [t_i, t_{i+1}] $, $ K(t) \approx \phi_{out}(t) $, and $ \rho(y,t) $ remains well approximated by Eq. (\ref{eq:rho_matching}).

Therefore, in the case treated here, Eq. (\ref{rate1}) becomes
\begin{equation}
r(t) = \frac{ \mid \beta c(t) + g(t) \mid }{\sqrt{2\pi\sigma^2(t)}} \exp \left( -\frac{ [\beta - F(t)]^2 }{2\sigma^2(t)} \right). 
\label{rate1_app}
\end{equation}

\begin{figure}
	\centering
	\includegraphics[width=0.8\textwidth]{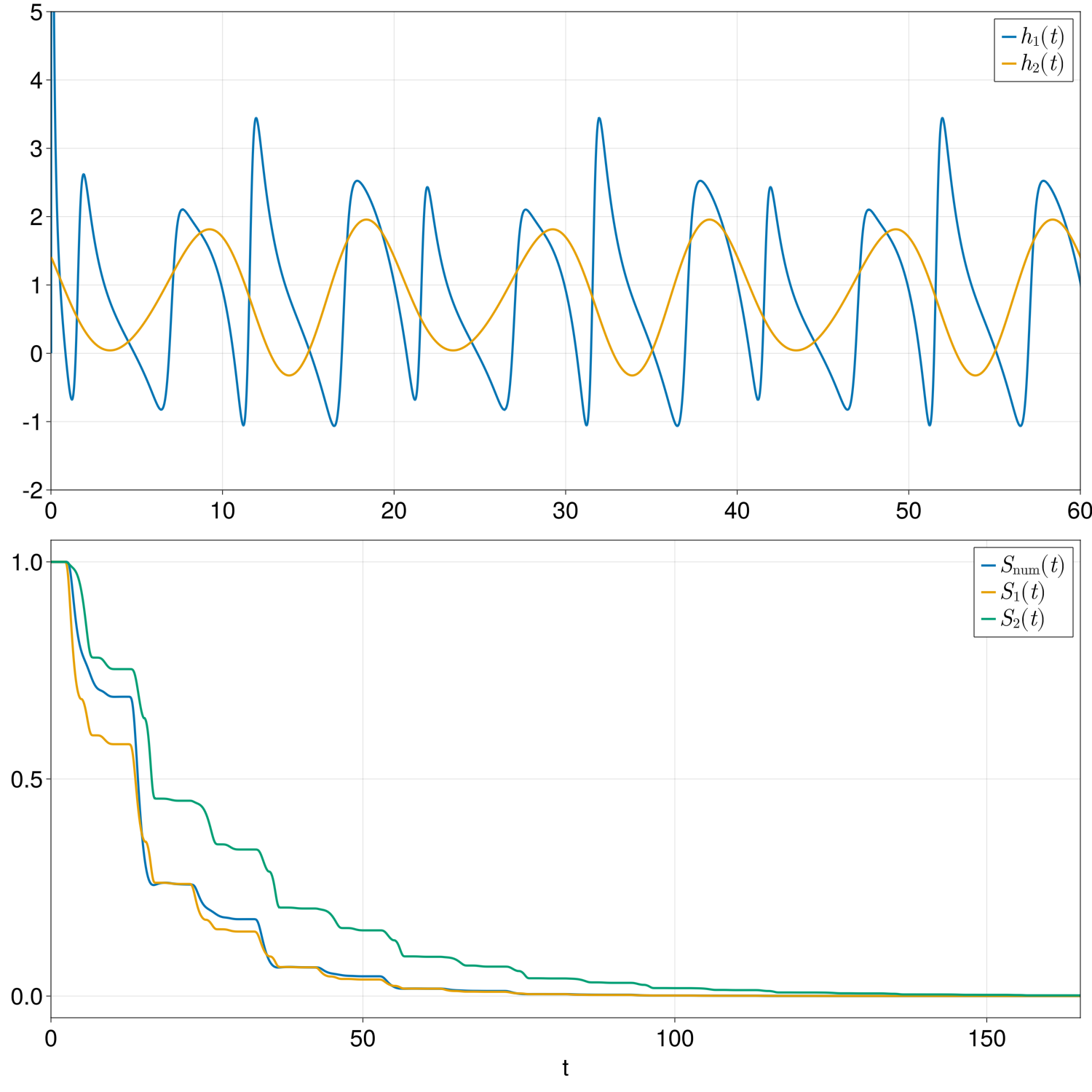}
	\caption{First panel: Comparison of \( h_1(t) \) and \( h_2(t) \) as derived in Sections \ref{sec3} and \ref{sec4} respectively, using the time dependent coefficients of Eq. (\ref{neg_params}). Second panel: Comparison between the survival probabilities computed numerically ($S_{num}(t)$: blue lines), and the approximate analytical results described in Section \ref{sec3} ($S_1(t)$: orange lines), and in Section \ref{sec4} ($S_2(t)$: green lines).}
\label{Fig5}
\end{figure}

As a concrete example, we explored the Langevin Eq. (\ref{lang}) using the following time dependent model parameters
\begin{equation}
\begin{split}
a(t) &= 1 - 0.7 \sin \left( \frac{1}{5} \pi t \right) + 0.7 \cos \left( \frac{1}{5} \pi t \right), \\
b(t) &= 1 - \sin \left( \frac{2}{5} \pi t \right), \text{~~~and}\\
f(t) &= 0.5 + 0.5 \sin \left( \frac{3}{10} \pi t \right) + 0.5 \cos \left( \frac{3}{10} \pi t \right).
\end{split}
\label{neg_params}
\end{equation}
These allow both \( h_1(t) \) and \( h_2(t) \) to take on negative values for a subset of time explored, as illustrated in the first panel of Figure (\ref{Fig5}).

In the second panel of Figure (\ref{Fig5}), we compare the approximate analytical results for the survival probabilities $S_1(t)$ and $S_2(t)$, obtained using the two different methods of Sections \ref{sec3} and \ref{sec4} respectively, with the numerical results, $S_{num}(t)$.  Clearly, the asymptotic methods of Section \ref{sec3} are superior.

\begin{figure}
	\centering
	\includegraphics[width=0.8\textwidth]{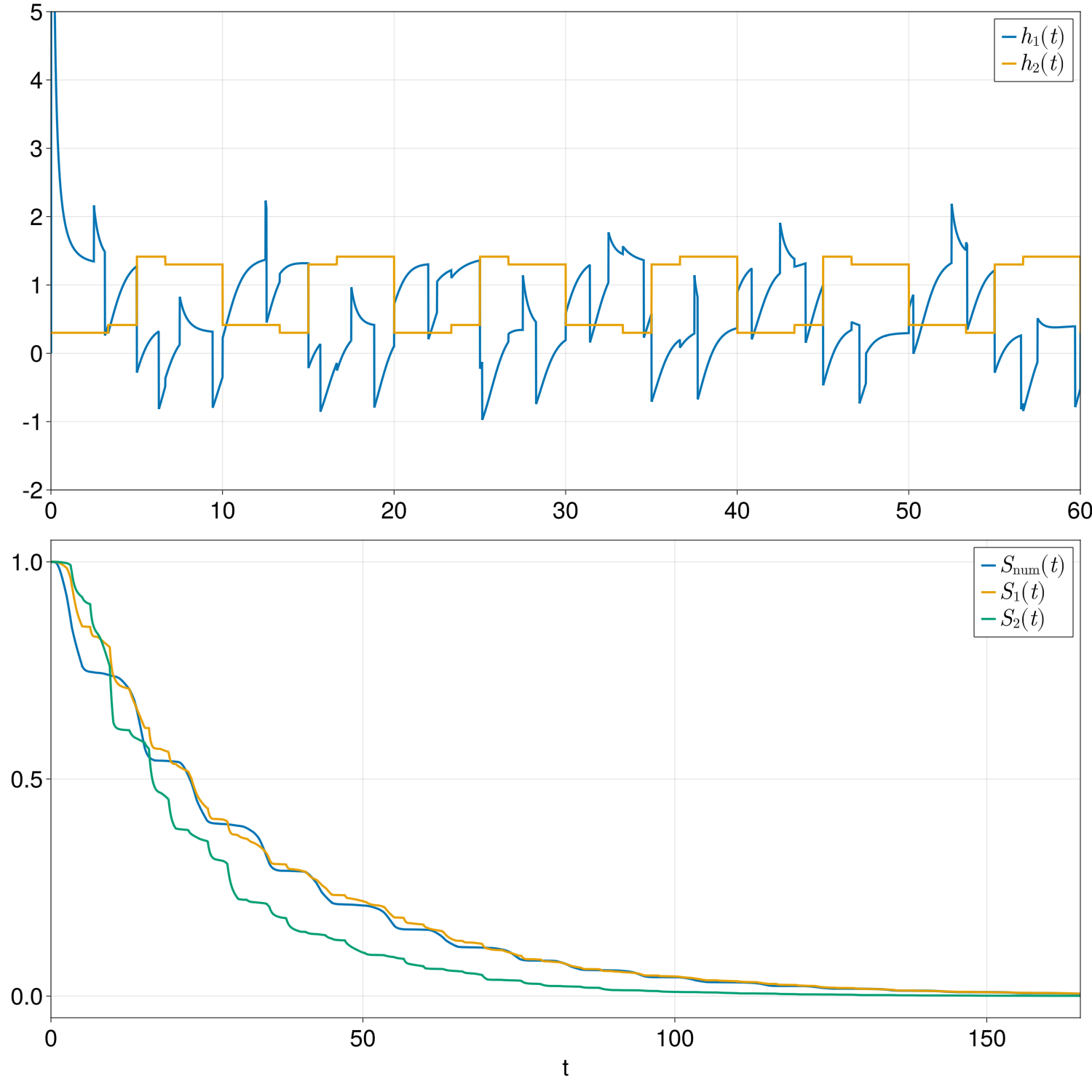}
	\caption{\textcolor{black}{Same as Figure (\ref{Fig5}), but using the time dependent coefficients of Eq. (\ref{SW_params}).}}
\label{Fig6}
\end{figure}

\textcolor{black}{
\section{Square Waves}\label{secA2b}
In this Appendix, we explore the case where square waves are used as the periodic functions for the time-dependent coefficients $a(t)$, $b(t)$, and $f(t)$, rather than the sinusoidal functions used previously.
We consider the Langevin Eq.~(\ref{lang}) with the following time-dependent model parameters:
\begin{equation}
\begin{split}
a(t) &= 1 - 0.5 \, \textrm{sign}\left[\sin\left(\frac{1}{5}\pi t\right)\right], \\
b(t) &= 1 - 0.2 \, \textrm{sign}\left[\sin\left(\frac{2}{5}\pi t\right)\right], \text{~~~and}\\
f(t) &= 0.5 + 0.2 \,\textrm{sign}\left[\sin\left(\frac{3}{10}\pi t\right)\right].
\end{split}
\label{SW_params}
\end{equation}
In the first panel of Figure (\ref{Fig6}) we plot \( h_1(t) \) and \( h_2(t) \) constructed using the time periodic coefficients of Eq. (\ref{SW_params}).
In the second panel of Figure (\ref{Fig6}), we compare the approximate analytical results for the survival probabilities $S_1(t)$ and $S_2(t)$, obtained using the two different methods of Sections \ref{sec3} and \ref{sec4} respectively, with the numerical results, $S_{num}(t)$.  As is the case throughout this paper, the asymptotic methods of Section \ref{sec3} are superior.
}

\section{Parameters used in Fig.~(\ref{Fig4})}\label{secA3}

Here we report the values of $\beta$ and $a(t), b(t)$ and $f(t)$ used to generate Fig.~(\ref{Fig4}). The coefficients have been randomly chosen with the same constraints described in Section \ref{sec4}.

\begin{equation}
\begin{split}
&\textrm{First\,row:}\\
&\beta=2.4,\\
&a(t)=1+0.69\sin(0.3t)-0.1\cos(0.3t),\\
&b(t)=1+0.68\sin(0.29t)-0.16\cos(0.29t),\\
&f(t)=0.17+0.02\sin(0.39t)+0.12\cos(0.39t).
\nonumber
\end{split}
\end{equation}
\begin{equation}
\begin{split}
&\textrm{Second\,row:}\\
&\beta=3.0,\\
&a(t)=1-0.64\sin(0.25t)+0.27\cos(0.25t),\\
&b(t)=1+0.7\sin(0.17t)-0.08\cos(0.17t),\\
&f(t)=0.02+0.01\sin(0.17t)+0.01\cos(0.17t).
\nonumber
\end{split}
\end{equation}
\begin{equation}
\begin{split}
&\textrm{Third\,row:}\\
&\beta=3.6,\\
&a(t)=1-0.5\sin(0.15t)-0.49\cos(0.15t),\\
&b(t)=1+0.56\sin(0.14t)-0.42\cos(0.14t),\\
&f(t)=0.33-0.08\sin(0.59t)+0.22\cos(0.59t).
\nonumber
\end{split}
\end{equation}
\begin{equation}
\begin{split}
&\textrm{Fourth\,row:}\\
&\beta=4.2,\\
&a(t)=1-0.45\sin(0.18t)-0.54\cos(0.18t),\\
&b(t)=1+0.56\sin(0.28t)-0.42\cos(0.28t),\\
&f(t)=0.03-0.01\sin(0.14t)-0.02\cos(0.14t).
\nonumber
\end{split}
\end{equation}

\end{appendices}





\end{document}